\documentstyle[11pt,epsf]{article}

\def\bea{\begin{array}}
\def\ena{\end{array}}
\def\bee{\begin{equation}}
\def\ene{\end{equation}}
\def\beqa{\begin{eqnarray}}
\def\enqa{\end{eqnarray}}

\def\t{\tau_{p\rightarrow e^+\pi^0}}

\newcommand{\ot}{\otimes}

\begin{document}

\title{Neutrino masses and baryogenesis in SO(10) unified theories}
\author{F. Buccella\thanks{Buccella@axpna1.na.infn.it}$~$,$~$
G. Mangano\thanks{Mangano@axpna1.na.infn.it} $~$ and $~$ L.
 Rosa\thanks{Rosa@axpna1.na.infn.it}\\ \\
Dipartimento di Scienze Fisiche, and INFN, Sezione di Napoli \\
Universit\`a di Napoli {\it Federico II}, \\ 
I-80125 Napoli, Italy} 

\date{}
\maketitle
\begin{abstract}
We report on some phenomenological implications of a class of unified
models based on $SO(10)$ gauge group, with intermediate symmetry
group containing $SU(2)_R$. Interesting predictions for neutrino masses
are discussed, which are relevant both for solar neutrino and
dark matter problems, as well as a model for the formation of the
baryon asymmetry of the universe required by primordial nucleosynthesis.

\end{abstract}

\newpage

\baselineskip20pt

\section{Introduction}

\vskip12pt

It is now over twenty years, when the $SU(5)$ model 
was proposed by Georgi and Glashow, that {\it unification programme} 
has been carried over, looking for a Gauge Unified Theory
(GUT) based on a larger symmetry simple group $G$ embedding the
standard one,
$G\supset G_{321}=SU(3)_c\otimes SU(2)_L\otimes U(1)_Y$, to which
it breaks down.

The most peculiar signature of these theories is the non
conservation of baryonic number which, in particular, allows for
proton instability.
For long time the comparison of the predicted proton lifetime with
the experimental lower bound has been the only crucial test that
GUT's should satisfy.
In this respect the minimal version of the $SU(5)$ model, in which 
Higgs scalars are classified in $24\oplus 5\oplus \bar{5}$, is
at variance with the very precise measurements of the three coupling
constants $\alpha_s$, $\alpha_L$ and $\alpha_Y$ of, respectively,
$SU(3)_c$, $S(2)_L$ and $U(1)_Y$ at the $Z^0$ mass scale. Actually
in 1990-1991 \cite{ambu,amfu} it became clear
that the three couplings do not match at a single point, and this
rules out the minimal $SU(5)$ model.

The big interplay between particle physics and cosmology, as well
as the increase in precision of cosmological measurements,
open the perspective of using the history of universe as a natural 
laboratory in which to check the predictions at high energies of
classes of unified theories. Items as the production of heavy
monopoles, the origin of baryonic matter-antimatter asymmetry or
the nature of dark matter in the galactic halos, provide a set of
additional constraints which are, sometimes, quite stringent.

In this paper we would like to account for the main features of a
class of models based on the $SO(10)$
gauge group. Interestingly, they are able to predict a value for
$\tau_{p\rightarrow e^{+}\pi ^0}$ in agreement with the experimental
lower limit, as well as neutrino masses of the order of magnitude
required to explain solar neutrino problem using MSW \cite{miwo},
and to give the status to $\tau$ neutrinos of interesting candidates
for the {\it hot} component of the dark matter.
Moreover, we will discuss how, within $SO(10)$ models, 
the three conditions necessary to have production of a baryon
asymmetry \cite {saka} appears to be naturally satisfied and a
value for the baryon to photon density in agreement with data on
primordial nucleosynthesis emerges.

\section{$SO(10)$ Gauge Theories}

\vskip12pt

SO(10) has been proposed \cite{geor} as a unifying gauge group with
three main motivations:

\begin{itemize}
\item[a)] models based on this gauge group are naturally anomaly free.
What appeared as a chance in SU(5) model, because of their
exact compensation due to the use of 10 and $\overline{5}$ to classify
fermions, is a general feature of all orthogonal
groups, with the only exception of $SO(6)$;
\item[b)]  the 10 and $\overline{5}$ representations 
of SU(5) are contained in
the 16 (spinorial) representation of SO(10) together with a SU(5)
singlet
with the quantum numbers of a $\overline{\nu }_L$;
\item[c)]  the 16 of SO(10) decomposes under $SU(4)\otimes SU(2)
\otimes SU(2)$,
the gauge group first introduced by Pati and Salam, into the $(4,2,1)
+(\overline{4},1,2)$ representation, which displays the
quark-lepton universality of weak interactions.
\end{itemize}

Quite recently other interesting features of the models have been
also investigated:

\begin{itemize}
\item[d)] the possibility, through the see-saw mechanism
\cite{gera}, to give cosmologically and astrophysically interesting
Majorana mass to neutrinos;
\item[e)] the explanation of baryon asymmetry of the universe 
through out-of-equilibrium decays of heavy Higgs bosons and of low 
energy sphaleron-like processes.
\end{itemize}

A study of the low dimensional Irreducible Representations (IR's)
of $SO(10)$ shows that their singlets (with the only exception of the
vector-spinor representation 144) have symmetry larger than $G_{321}$. 
Therefore, at least two of them are necessary to drive the symmetry
breaking process of $SO(10)$ down to $G_{321}$, via an intermediate
symmetry stage with an unbroken subgroup $G '$, whose energy scale 
we will denote with $M_R$.
Beyond $SU(5)$ (or $SU(5)\ot U(1)$), there are
four intriguing cases which correspond to several $G'$. 
They are listed in Table I togheter with the direction of the 
minimum of Higgs potential.
\vskip2truecm
{
\small
\vskip15pt

\begin{center}
{Table I}

\vskip10pt

\begin{tabular}{|c|c|c|c|}
\hline
&  &  & \\ 
& $G^{\prime }$ & Higgs direction & IR \\ 
&  &  & \\ \hline
&  &  & \\ 
A  & $SU(4)_{PS}\otimes SU(2)_L\otimes SU(2)_R\times D$ &
$\omega _L=\frac 2{
\sqrt{60}}\left( \omega _{11}+\ldots +\omega _{66}\right)
-\frac 3{\sqrt{60}
}\left( \omega _{77}+\ldots \omega _{00}\right) $ & 54 \\ 
&  &  & \\ \hline
&  &  & \\ 
B & $SU(3)_c\otimes SU(2)_L\otimes SU(2)_R\otimes U(1)_{B-L}\times D$
& $\Phi _L=
\frac{\Phi _{1234}+\Phi _{1256}+\Phi _{3456}}{\sqrt{3}}$ & 210 \\ 
&  &  & \\ \hline
&  &  & \\
C & $SU(4)_{PS}\otimes SU(2)_L\otimes SU(2)_R$ & $\Phi _T=\Phi _{7890}$
& 210 \\
&  &  & \\ \hline
&  &  & \\ 
D & $SU(3)_c\otimes SU(2)_L\otimes SU(2)_R\otimes U(1)_{B-L}$ & $\Phi
(\theta
)=\cos \theta \Phi _L+\sin \theta \Phi _T$ & 210 \\ 
&  &  & \\ \hline
\end{tabular}
\end{center}
$\omega _{ab}$ is a second-rank traceless symmetric tensor;
$\Phi _{abcd}$ is a fourth-rank antisymmetric tensor, and the
indices 1...6 correspond to $SO(6)\sim SU(4)_{PS}$, whereas 7...0
correspond
to $SO(4)\sim SU(2)_L\otimes SU(2)_R$.
}
\bigskip

For the models in Table I it is possible to show that, at one loop
approximation, and within the Extended Survival Hypothesis (ESH),
i.e. by only considering scalars in the renormalization group
equations (RGE) which are required to drive symmetry breaking
at $M_R$ and at the electroweak scale, the first
breaking scale $M_X$ corresponding to $SO(10) \rightarrow G^{\prime}$
is not higher than the meeting point of $\alpha _3$ and $\alpha _2$
in the SU(5) minimal model \cite{bumi}. In particular, with two
doublets of scalars in the
10 (to avoid the prediction $m_t=m_b$), one can get the
inequality
\begin{equation}
M_X\leq M_{Z}\exp {\ \frac \pi
{2\alpha {(M_{Z})}}\left( \sin ^2\theta
_W(M_{Z})-\frac \alpha {\alpha _s}{(M_{Z})}\right) }
\end{equation}
Since the lepto-quarks responsible for proton decay take mass at 
$M_X$, this inequality traduces into an upper limit for exclusive
processes lifetime, since $\tau_p \propto M_X^4$.
A more restrictive bound, by a factor of about $\frac{1}{3}$, can be
obtained using a two-loop approximation \cite{acam}.

It is also worth reminding that the intermediate scale $M_R$ at which 
$G^{\prime}$ breakes down to the standard group $G_{321}$ is
connected, via the see-saw
mechanism \cite{gera}, to Majorana mass for $\bar{\nu}_L$.
This allows for quite interesting 
prediction for $\nu _\tau$ and $\nu _\mu $ masses. We will 
come back to this point later on. For the value of $M_R$,
at one loop and still using ESH, it was possible to find an upper
limit for $M_R$ 
\begin{equation}
M_R\leq M_Z\exp {{\frac \pi {6\alpha (M_Z)}}\left[
\frac 32-3\sin ^2{\theta
_W(M_Z)}-{\frac \alpha {\alpha _S}}(M_Z)\right] },
\end{equation}
where the equality holds only for the model with $G^{\prime }\supset
SU(4)_{PS}\times D$, whose prediction for $M_X$, however, is too 
small when compared with experimental lower limit.

Because of the richness of the mass spectrum of $SO(10)$ models,
it may seem
that ESH is a too restrictive assumption. Actually this observation
led Dixit
and Sher to claim that huge uncertainties are introduced in the
$SO(10)$ predictions if ESH is removed \cite{dish}. However,
as it has been explicitly shown in a recent paper \cite{acam}, by
carefully studying the mass spectrum, it is possible to
deduce rather restrictive conditions on the contributions of the
scalars to RGE.
In this way, by requiring for $M_X$ a value sufficiently high to be 
in agreement with the lower limit on
$\tau _{p\rightarrow e^{+}\pi ^0}$, it is possible
to find upper limits on $M_R$, which correspond to lower
limits on the masses of the (almost) left-handed
neutrinos.

\section{Prediction on symmetry breaking scales}

To obtain the values of $M_X$ and $M_R$ it is necessary to study the 
mass spectrum of scalars that,
according to the decoupling theorem of Appelquist-Carrazzone, 
would contribute to RGE. In the following we quote the results of a 
detailed study performed
on the mass spectrum characterizing the various models of Table I.
(for a complete analysis see \cite{acam}). The mass of the multiplet
$(d_1,...,d_n)$, transforming as a $d_i$ dimensional IR under the
factor $G_i \subset G'$, will be denoted with $m(d_1,...,d_n)$

\subsection{$SU(4)\otimes SU(2)\otimes SU(2)\times D$}

In the following we will use the values for $sin^2 \theta_W$,
$\alpha_s$ and $\alpha$ at the $M_Z$ mass energy scale \cite{lang}
\beqa
\sin^2 \theta_W(M_Z) &=& 0.2315\pm 0.0002\pm 0.0003 \\
\alpha_s \left( M_Z \right) &=& 0.123\pm 0.004\pm 0.002 \\
\alpha \left( M_Z \right) &=& 1/(127.9\pm 0.2)
\enqa
where the second error on the first two results is due to the
uncertainty on the Higgs mass.

For this model we find, at one loop, 
\begin{equation}
M_R=M_Z\exp {B}=M_{SU(5)} =5.3\cdot10^{13}~GeV \label{eq:rlor}
\end{equation}
with $B~\equiv ~{\frac \pi {11\alpha (M_Z)}}\left(
{\frac 32}-3\sin ^2{%
\theta _W(M_Z)}-{\frac \alpha {\alpha _S}}(M_Z)\right) $ .

Thus, in this case, it is not possible to increase the value of $M_R$.
If one tries to get $M_X\geq 3.2\cdot 10^{15}GeV$ it is necessary to
take into account the request of having minimum in the
$SU(4)\otimes SU(2) \otimes SU(2)\times D$ direction, which gives,
still at one loop,
\begin{equation}
M_X < M_Z~e^{{\frac 1{21}}(11A+9B+3\ln {1.35})}=  
 (2.46\cdot 10^{15})~GeV
\end{equation}
with 
$A~\equiv ~{\frac{6\pi }{11\alpha (M_Z)}}\left( \sin ^2{\theta _W(M_Z)}-%
{\frac \alpha {\alpha _S}}(M_Z)\right) $.

The same study can be performed at two loops in RGE. In this case we
find
(with all the multiplets at the scale $M_R$ but the (1,3,3) one,
for which we have $m(1,3,3)\leq 1.35~M_R$): 
\begin{equation}
\begin{array}{ll}
M_R^{(2)}=4.9\cdot 10^{13}~GeV, & M_X^{(2)}=0.74\cdot 10^{15}~GeV
\end{array}
\end{equation}
$M_X^{}$\ is too small (about two standard deviations) to comply with
the lower limit on proton lifetime.

\subsection{$G^{\prime}\equiv SU(4)_c\otimes SU(2)_L\otimes SU(2)_R$}

In this case, one obtains for $M_R$: 
\begin{equation}
\ln {\frac{M_R}{M_Z}}=B-\frac 5{44}\Biggr[ 3\ln {\frac{m(15,3,1)}
{m(15,1,3)}}
+4\ln {\frac{m(10,3,1)}{m(\overline{10},1,3)}}\Biggr]  \label{eq:btiz}
\end{equation}

The ESH would imply $m(1,2,2)_{10}\sim M_Z$,
$m(\overline{10},1,3)\sim M_R$
and all the other scalars with mass $\sim M_X$. One would obtain for
$M_R$ (this time $M_X$ is consistent with the experimental limit)
\begin{equation}
M_R^{}=2.3\cdot 10^{11}~GeV, ~M_X=6.2 \cdot 10^{15}~GeV
\end{equation}

To establish how strongly this result depend on ESH, we look for the
highest value for $M_R$ consistent with a sufficiently
high value for $M_X$ ($M_X^{(2)}\geq 3.2\cdot10^{15}~GeV$).

From eqs. (\ref{eq:btiz}),
by taking $m(10,3,1)\geq M_R
$, in order to have the highest $M_R$ and the absolute minimum in the
$G_{422}$-invariant direction one has: 
\begin{eqnarray}
m(10,3,1)_{210} &=&m(\overline{10},1,3)_{210}, \label{1031}\\
m^2(15,3,1)_{210} &=&\frac{2-\sqrt{3}}{2+\sqrt{3}}m^2(15,1,3)_{210},
\label{1531} \\
\frac{m^2(15,3,1)_{210}m^2(15,1,3)_{210}}{m^4(15,1,1)_{210}} &\leq &9
\label{1511}.
\end{eqnarray}
From eqs. (\ref{1031}) (\ref{1531}) and (\ref{1511}), one
gets the following inequality: 
\begin{equation}
M_R \leq 3^{\frac{1}{14}} M_Z \left(\frac{M_Z}
{M_X}\right)^{\frac{10}{7}}
e^{\frac{\pi}{14}\left(\frac{3}{\alpha(M_Z)}-\frac{8}{\alpha_s(M_Z)}
\right)}
\leq 4.87 \cdot 10^{13}~GeV,
\end{equation}
the last one coming from $M_X\geq3.2\cdot10^{15}~GeV$.

The highest value for $M_R$ is found by taking $(1,2,2)_{10}$ at the
scale 
$M_Z$, the multiplets $(10,2,2)_{210},~(6,1,1)_{10}$ and all the states
of
the 126 at the scale $M_R$ and all the other states of the 210 at
the scale 
$M_X$: in such conditions one finds: 
\begin{equation}
\begin{array}{ll}
M^{(2)}_R=1.6\cdot10^{13}~GeV, & M^{(2)}_X=3.2\cdot10^{15} ~GeV
\end{array}
\end{equation}

\subsection{$G^{\prime}\equiv SU(3)_c\otimes SU(2)_L\otimes SU(2)_R
\otimes U(1)_{B-L}\times D$}

In this case, by keeping into account the $SU(2)_L\leftrightarrow
SU(2)_R$
symmetry above $M_R$, ESH would imply $m(1,2,2,0)_{10}\sim M_Z$, 
$m(1,3,1,-2)_{126}\sim M_R$ and all the other multiplets at the scale
$M_X$.
One would get at two loops 
\begin{equation}
\begin{array}{ll}
M_R^{(2)}=5.3\cdot 10^{10}~GeV, & M_X^{(2)}=1.98\cdot 10^{15}~GeV
\end{array}
\end{equation}
a too small value for $\tau _{p\rightarrow e^{+}\pi ^0}^{(2)}$.
In considering the contribution of the other scalars, the constraints
on the mass spectrum, which follow from the
requirement that
the absolute minimum falls in the desired direction, implies the
following inequalities for the masses 
\begin{eqnarray}
\frac{m(8,3,1,0)_{210}}{m(3,3,1,4/3)_{210}}>\sqrt{\frac{37}{14}} &,
&\frac{%
m(6,2,2,2/3)_{210}}{m(1,2,2,2)_{210}} >\frac{1}{\sqrt{7}}, \\
1<\frac{m(8,3,1,0)_{210}}{m(6,2,2,2/3)_{210}}<\frac{2}{\sqrt{3}}& ,
& 1\leq
\frac{m(3,3,1,-2/3)_{126}}{m(3,2,2,4/3)_{126}}\leq 2 .  \label{eq:buc}
\end{eqnarray}

In this way it is found: 
\begin{eqnarray}
M_R &\leq& \left(\frac{2\cdot 7^{\frac{5}{3}}}{37}\right)^{\frac{3}
{31}}
\frac{M_{SU(5)}^{\frac{44}{31}}}{M_X^{\frac{13}{31}}}
\label{eq:mrfir} \\
M_R &\leq& M_Z^{\frac{5}{6}}M_X^{\frac{1}{6}} e^{{\frac{\pi}
{\alpha(M_Z)}}%
\left({\frac{1}{4}}-\sin^2{\theta_W(M_Z)}+\frac{1}{3}{\frac{\alpha}
{\alpha_s}%
}(M_Z)\right)}  \label{eq:mrsec}
\end{eqnarray}
where the equality holds only in the extreme case with
$m(8,1,1,0)_{210}$,
$%
m(6,3,1,2/3)_{126}$ and $m(8,2,2,0)_{126}$ at the scale $M_R$ and $%
m(3,2,2,4/3)_{126}$ at the scale $M_X$ for eq. (\ref{eq:mrfir}), and $%
m(3,2,2,-2/3)_{210}$, $m(3,3,1,-2/3)_{126}$ at the scale $M_R$ and $%
m(1,2,2,2)_{210}$, $m(8,1,1,0)_{210}$ and $m(3,1,1,-2/3)_{126}$ at the
scale
$M_X$ for eq. (\ref{eq:mrsec}). The two requirements may not be
satisfied at
the same time, since they imply a different scale for
$m(8,1,1,0)_{210}$ and
disagree with eq. (\ref{eq:buc}), which implies that the concerned
masses are at the same scale.

By eliminating $M_X$ in eqs. (\ref{eq:mrfir}) and (\ref{eq:mrsec})
one finds the inequality 
\begin{equation}
M_R<3.8\cdot10^{11}~GeV.  \label{eq:buc1}
\end{equation}

It would be possible, of course, to get a lower bound for $M_R$
since the one
just written has been obtained by multiplying inequalities which
cannot be
both equalities. So it is not surprising that, by looking for the
highest
value for $M_R$ consistent with $M_X^{(2)}\geq 3.2\cdot 10^{15}~GeV$
and
with the constraints on the spectrum following, we find 
\begin{equation}
M_R^{(2)}=2.7\cdot 10^{10}~GeV.
\end{equation}

\subsection{$G^{\prime}\equiv SU(3)_c\otimes SU(2)_L\otimes SU(2)_R
\otimes U(1)_{B-L}$}

In the ESH limit 
\begin{equation}
\begin{array}{ll}
M_R^{(2)}=5\cdot 10^9GeV & M_X^{(2)}=1.3\cdot 10^{16}GeV.
\end{array}
\end{equation}

Due to the complexity of the conditions, we have not been able to
deduce, as
in the previous cases, intriguing inequalities for the one-loop
equations,
and a lenghty numerical analysis has been needed to get the highest
value
for $M_R$ \cite{acam}. As a result, we found with the scalars of
the $%
126\oplus \overline{126}$ at the scale $M_R$ and with the scalars
of the 210 at the scale $M_X$

\begin{equation}
M_R^{(2)} \leq 0.48\cdot 10^{11}~GeV
\end{equation}

\section{Baryogenesis}

As shown by Sakharov \cite{saka}, three conditions should be satisfied
to produce the baryon asymmetry, which is required in order to explain
the result on baryon to photon density $\eta_B$ in the universe,
coming from primordial nucleosynthesis data, $\eta_B \sim (3 \div 4)~
10^{-10}$
\begin{itemize}
\item[1)] baryon number ($B$) violating interactions;
\item[2)] $C$ and $CP$ violation;
\item[3)] non equilibrium conditions.
\end{itemize}
Because of 1), it was soon realised that GUT theories may be the
natural framework for baryon asymmetry generation. In the {\it
standard} scenario this asymmetry is produced by out of equilibrium
decays of heavy Higgs scalars or gauge bosons into light fermions
\cite{kolb}.
However it was pointed out by several authors \cite{shap}
that anomalous $B+L$ violating processes ($L$ is lepton number)
mediated by sphaleronic
$SU(2)_L \otimes U(1)_Y$ gauge and Higgs configurations at low energy
scale can almost
completely wash out any asymmetry in $B$ or $L$ produced at GUT scales.
Nevertheless, it is worth observing that these kind of effects
cannot affect an asymmetry produced for $B-L$, since the
corresponding current is anomaly free. A possible scenario for the
production of $\eta_B$ is therefore based on the idea that an asymmetry
in this quantum number is produced at GUT scales and then it is
eventually transformed into $B$ and $L$ asymmetries at low scales
via the {\it shuffling} effect of sphaleronic configurations.

This possibility has been studied in the framework of $SO(10)$ models
in \cite{buma}, to which we refer for details.
Notice that for minimal $SU(5)$ model, $B-L$ represents
an accidental global symmetry and, therefore, no corresponding asymmetry can
be produced in this case.

An interesting feature of $SO(10)$ models is the possibility to define
a charge conjugation operator $C$ whose
corresponding symmetry remains unbroken even after $M_X$ if $G'$ has
rank 5 \cite{hase}. Notice that all models considered in the previous
sections are just of this kind.
It follows that any asymmetry in $C$-odd quantum
number, as $B-L$, can only be produced after $C$ is 
broken, i.e. when $G'$ is spontaneosuly broken down to the standard
model at $M_R$. Interestingly enough, this scale is also the one
at which $B-L$ is no more a symmetry of the vacuum,
what allows, if $CP$ violating effects are also present, for the
production of microscopic asymmetries $\delta_{B-L}$
in decay or scattering processes.
In particular we have considered decays of heavy Higgs bosons $\Phi$
in channels containing massive $\bar{\nu}_L$. Tipically one expects
for $\delta_{B-L}$ values of the order
\begin{equation}
\delta_{B-L} \sim h^2 \frac{M_{\nu}^2}{M_{\Phi}^2} \epsilon_{CP}
\label{order}
\end{equation}
where $h$ denotes the order of magnitude of the Yukawa couplings
which are involved in the process, $M_{\nu}$
are the Majorana neutrino masses, $M_{\Phi}$ is the mass of the
decaying particle
and finally $\epsilon_{CP}$ parametrizes the magnitude of $CP$
violation. However, this microscopic asymmetry traduces into a
macroscopic one only if decay processes occur out of equilibrium.
If this is not the case, in fact, inverse decay processes and
scattering mediated by virtual propagating $\Phi$, would produce
an opposite value for $\delta_{B-L}$. Actually, from a
thermodynamical point of view, no asymmetry corresponding to non
exactly conserved quantum numbers can be generated in equilibrium
conditions, since maximal entropy is obtained for vanishing
corresponding chemical potentials.

We remind that a quite accurate indications of whether a decay process
occurs in equilibrium or not is to compare the corresponding width
$\Gamma$ to the current value of the Hubble parameter $H$ when the
decaying massive particle becomes non-relativistic ($T \sim M_{\Phi}$).
In particular
out of equilibrium conditions corresponds to $\Gamma \leq H$.
In radiation dominated epoch, it holds $H \sim \sqrt{g}~ T^2
/M_{Pl}$ with $T$ the temperature, $m_{Pl}$ the Planck mass and $g$ the
number of effective degrees of freedom ($g \sim 100$ at GUT scales).
A more
sophisticated analysis, using Boltzmann equations gives, in general,
similar
results \cite{kolb}.
The most natural candidates to produce a $B-L$ asymmetry are the
multiplets in the 126 IR which takes mass at $M_R$ and can decay in
a $\bar{\nu}_L$ and a fermion $f$ pairs.
However, because of the quite low value
of $M_R$ compared with Planck mass, these decays occurs in equilibrium
since
\begin{equation}
\Gamma (126 \rightarrow \bar{\nu}_L f) = \frac{h^2}{32 \pi} M_R  >
H(T=M_R) \sim 10 \frac{M_R^2}{M_{Pl}}
\end{equation}
unless $h$ is chosen unnaturally small.

For this reason, it seems that more suitable candidates for the $\Phi$
we are looking for,
should have larger mass than $M_R$, taken at the first
symmetry breaking stage at $M_X$. In addition they
also have to be weakly coupled in order to survive down to $M_R$, at
which $B-L$ violating processes become possible. A natural candidate
satisfying the above conditions are the Higgs bosons in the 210
IR. They cannot be coupled to pair of fermions at tree level because
210 IR is not contained in the product 16 $\times$ 16. Moreover
effective terms in the lagrangian at higher order in perturbation
theory are also absent if the intermediate group $G'$ has rank 5.
Since the fast decay channel in fermions pairs is forbidden, it has been
shown in \cite{buma}, for the case $G' = SU(3)_C \otimes SU(2)_L
\otimes SU(2)_R \otimes U(1)_{B-L}$,
that the most efficient channel is $210 \rightarrow
126~ f~f$, whose rate is sufficiently small to have an {\it out of
equilibrium} overabundant population of these bosons at $M_R$\footnote{
To evaluate the residual density as a function of temperature it is also
necessary to take into account annihilation processes. A study of these
effects has been considered in details in \cite{buma}, showing that they
are {\it frozen out} and unable to strongly reduce the 210 population
at $M_R$ if their mass is larger than $10^{12}~GeV$, as it can be
expected since they take mass at $M_X \geq 3.2 \cdot 10^{15} ~GeV$.}.
When these decays eventually occur, for temperature below this scale,
the particular channels $210 \rightarrow 126~\bar{\nu}_L~ f$ produce
a microscopic $B-L$ asymmetry $\delta_{B-L}$ of the order of magnitude
reported in (\ref{order}). The macroscopic value for $B-L$ asymmetry,
$\Delta(B-L)$, in particular the ratio $\Delta(B-L) /s$, where $s$
is the specific entropy,
can be evaluated by considering the dilution effect due to
entropy release of decaying heavy bosons to radiation
degrees of freedom \cite{kolb}.
\begin{equation}
\frac{\Delta(B-L)}{s} \sim h_{10}^2 \frac{M_{\nu}^2}{M_{\Phi}^2}
\epsilon_{CP} \frac{T_{RH}}{M_{\Phi}}
\end{equation}
where $h_{10}$ and $h_{126}$ are the Yukawa coupling of 10 and 126 IR
to fermions, respectively and
$T_{RH}$ is the reheating temperature. This value,
due to sphaleronic effects, traduces into a corresponding value for
baryon asymmetry $\Delta B \sim \Delta (B-L)/3$, which for $M_{\Phi}$
in the range $10^{12} \div 10^{14}~GeV$ gives a result in agreement
with the experimental value coming from the results on primordial
nucleosynthesis of light nuclei, provided the $CP$ parameter
$\epsilon_{CP}$ is larger than $10^{-2}$, which sounds quite reasonable.

\section{Conclusions}

In TABLE II and III we have reported the result on $M_X$ and $M_R$
for the models discussed in the previous sections, while in Figure 1
are shown the curves of level, in the plane $sin^2\theta_W - \alpha_s$
corresponding to several values of $M_R$.\\
The results of the analysis can be summarized as follows:
\begin{itemize}
\item[$\bullet$] In case $A$ of Table I the resulting value of $M_X$,
both with
and without the ESH hypothesis, is too small in order to comply with
the
experimental limit on proton lifetime, so this possibility is ruled
out from the present determinations of $sin^2 \theta_W$ and $\alpha_s$;
\item[$\bullet$] case $B$ gives a value for $M_X$ which is in agreement
with
$\t$ only if ESH is released. Even in this case, the upper limit on
$M_R$ gives, via the see-saw mechanism, a lower limit on $\nu_{\tau}$
mass, namely $m_{\nu_{\tau}} \geq 80 \div 140 ~eV$,
depending on the values chosen for $sin^2 \theta_W$ and $\alpha_s$,
which is too high if compared with cosmological constraints on
the age of the universe,
\item[$\bullet$] in case $C$, in the ESH limit, one has
$m_{\nu_{\tau}} \sim 15 ~ eV$, which 
is in agreement with the idea that $\nu_{\tau}$ gives
a relevant contribution to the $hot$ component of dark matter. By
releasing ESH one gets instead
the lower limit
$m_{\nu_{\tau}} \geq 0.2 \div 0.3 ~eV$ and
$m_{\nu_{\mu}} \geq (1.8 \div 2.2) \cdot 10^{-5}~eV$;
\item[$\bullet$] for case $D$, the ESH analysis gives too high values
for neutrino masses, while the lower limit one gets
on $m_{\nu_{\mu}}$ in the general
case, namely
$m_{\nu_{\mu}} \geq (2 \div 6) \cdot 10^{-3} ~eV$, is
just of the right order of magnitude to allow for a solution of solar
neutrino problems in the framework of MSW model. The results on
$\tau$ neutrino, $m_{\nu_{\tau}} \geq 25 \div 75 ~eV$, are quite large,
though the smallest one is still marginally compatible with analysis
on dark matter composition and perturbation spectra. Moreover,
in this case, we have shown how baryon asymmetry can be efficiently
produced by combining $B-L$ violating processes at high scales and
sphaleronic
effects at low scales, with a result for $\eta_B$ which is in
agreement with data from primordial nucleosynthesis.
\end{itemize}
It is worth stressing, to conclude, the strong uncertainties on these
predictions coming from the corresponding ones on the gauge couplings
at $M_Z$, as can be seen from Figure 1.
\pagebreak

\begin{center}

Table II

{\footnotesize

\bigskip

\begin{tabular}{|c|c|c|} \hline
 & & \\
& & upper limit for $M_R$ \\
First-Loop (ESH)  & Second-Loop (ESH)   &
with $M_X\geq3.2\cdot10^{15}GeV$ \\ 
& & and without ESH \\ \hline
\multicolumn{3}{|c|}{ $SU(4)\ot SU(2)\ot SU(2)\times D $} \\ \hline
& & \\
${M_X}={M_Z}\exp\left[\frac{\pi(.78+1.08
\sin^2\theta_W-3.16\frac{\alpha}{\alpha_s})}{11\alpha}
\right]=1.3\cdot10^{15} $& 
$\frac{M_X}{3.2\cdot10^{15}GeV}=0.21$ 
& $\frac{M_X}{3.2\cdot10^{15}GeV}\leq0.23$\\
& & \\ \hline
& & \\
${M_R}={M_Z}\exp\left[\frac{\pi (1.5-3\sin^2\theta_W
-\frac{\alpha}{\alpha_s})}{11\alpha}
\right]=5.4\cdot10^{13}$ & 
$\frac{M_R}{10^{11}GeV}=490 $ & \\
& & \\ \hline
\multicolumn{3}{|c|}{ $SU(3)\ot SU(2)\ot SU(2)\ot U(1)\times D $} \\ \hline
& & \\
${M_X}={M_Z}\exp\left[\frac{\pi(3+18\sin^2\theta_W-
26\frac{\alpha}{\alpha_s})}{70\alpha_s(M_Z)}
\right]=5.1\cdot10^{15}$ & 
$\frac{M_X}{3.2\cdot10^{15}GeV}= 0.62$ &  \\
& & \\ \hline
& & \\
${M_R}=M_Z\exp\left[\frac{\pi(3-12\sin^2\theta_W
+4\frac{\alpha}{\alpha_s})}{10\alpha_s(M_Z)}
\right]=1.8\cdot10^{10}$ & $ 
\frac{M_R}{10^{11}GeV}=0.53 $ & $\frac{M_R}{10^{11}GeV}\leq0.27$ \\
& & \\ \hline
\multicolumn{3}{|c|}{ $SU(4)\ot SU(2)\ot SU(2)$} \\ \hline
& & \\
${M_X}=M_Z\exp\left[\frac{\pi(1.5+
15\sin^2\theta_W-19\frac{\alpha}{\alpha_s})}{47\alpha_s(M_Z)}
\right]=8.7\cdot10^{15}$ & 
$\frac{M_X}{3.2\cdot10^{15}GeV}=1.93$ &  \\
& & \\ \hline
& & \\
${M_R}=M_Z\exp\left[\frac{\pi(10.5-
36\sin^2\theta_W+8\frac{\alpha}{\alpha_s})}{47\alpha_s(M_Z)}
\right]=7.7\cdot10^{11}$ & 
$ \frac{M_R}{10^{11}GeV}=2.3 $ & $\frac{M_R}{10^{11}GeV}\leq163$ \\
& & \\ \hline
\multicolumn{3}{|c|}{ $SU(3)\ot SU(2)\ot SU(2)\ot U(1) $} \\ \hline
& & \\
${M_X}=M_Z\exp\left[\frac{\pi(\sin^2\theta_W-
\frac{\alpha}{\alpha_s})}{2\alpha_s(M_Z)}
\right]=4.1\cdot10^{16}$ & 
$\frac{M_X}{3.2\cdot10^{15}GeV}=4.06 $ &  \\
& & \\ \hline
& & \\
${M_R}=M_Z\exp\left[\frac{\pi(12-51\sin^2\theta_W
+19\frac{\alpha}{\alpha_s})}{34\alpha_s(M_Z)}
\right]=1.4\cdot10^{9}$ & $ 
\frac{M_R}{10^{11}GeV}=0.05 $ & $\frac{M_R}{10^{11}GeV}\leq0.48$ \\
& & \\ \hline
\end{tabular}

}

\bigskip
\end{center}
The values of $M_X$ and $M_R$ in the case $\sin^2{\theta_W}=
0.2315\pm 0.0002$ and $\alpha_s=0.123 \pm 0.004$. 
When $M_X$ is less than
the lower limit we omit to write the upper limit on $M_R$.

\pagebreak

\begin{center}

Table III

{\footnotesize

\bigskip

\begin{tabular}{|c|c|c|} \hline
 & & \\
& & upper limit for $M_R$ \\
First-Loop (ESH)  & Second-Loop (ESH)   &
with $M_X\geq3.2\cdot10^{15}GeV$ \\ 
& & and without ESH \\ \hline
\multicolumn{3}{|c|}{ $SU(4)\ot SU(2)\ot SU(2)\times D $} \\ \hline
& & \\  
$8.5\cdot10^{14}GeV $& 
$\frac{M_X}{3.2\cdot10^{15}GeV}=0.14$ & 
$\frac{M_X}{3.2\cdot10^{15}GeV}\leq0.15$ \\
& & \\ \hline
& & \\
$5.5\cdot10^{13}$ & 
$\frac{M_R}{10^{11}GeV}=500$ & \\
& & \\ \hline
\multicolumn{3}{|c|}{ $SU(3)\ot SU(2)\ot SU(2)\ot U(1)\times D $}
\\ \hline
& & \\
$2.7\cdot10^{15}$ & $\frac{M_X}{3.2\cdot10^{15}GeV}=0.34$ 
&  \\
& & \\ \hline
& & \\
$5.9\cdot10^{10}$GeV & $\frac{M_R}{10^{11}GeV}=1.6 $ & 
$\frac{M_R}{10^{11}GeV}\leq0.48$  \\
& & \\ \hline
\multicolumn{3}{|c|}{ $SU(4)\ot SU(2)\ot SU(2)$} \\ \hline
& & \\
$4.3\cdot10^{15}$ & $\frac{M_X}{3.2\cdot10^{15}GeV}=0.91$ 
 & \\
& & \\ \hline
& & \\
$1.6\cdot10^{11}GeV$ & 
$ \frac{M_R}{10^{11}GeV}=5.6  $ & $\frac{M_R}{10^{11}GeV}\leq130$ \\
& & \\ \hline
\multicolumn{3}{|c|}{ $SU(3)\ot SU(2)\ot SU(2)\ot U(1) $} \\ \hline
& & \\
$1.6\cdot10^{16}$ & $\frac{M_X}{3.2\cdot10^{15}GeV}=1.69$ 
 & \\
& & \\ \hline
& & \\
$6.5\cdot10^{9}GeV$ & $\frac{M_R}{10^{11}GeV}=0.21$  &
$\frac{M_R}{10^{11}GeV}\leq1.5$  \\
& & \\ \hline
\end{tabular}

}

\bigskip
\end{center}
The same as Table II but with $\sin^2{\theta_W}=0.2302
\pm0.0005$ and $\alpha_s=0.117\pm0.008$

\newpage


\epsfysize=5.2in \centerline{\epsfbox{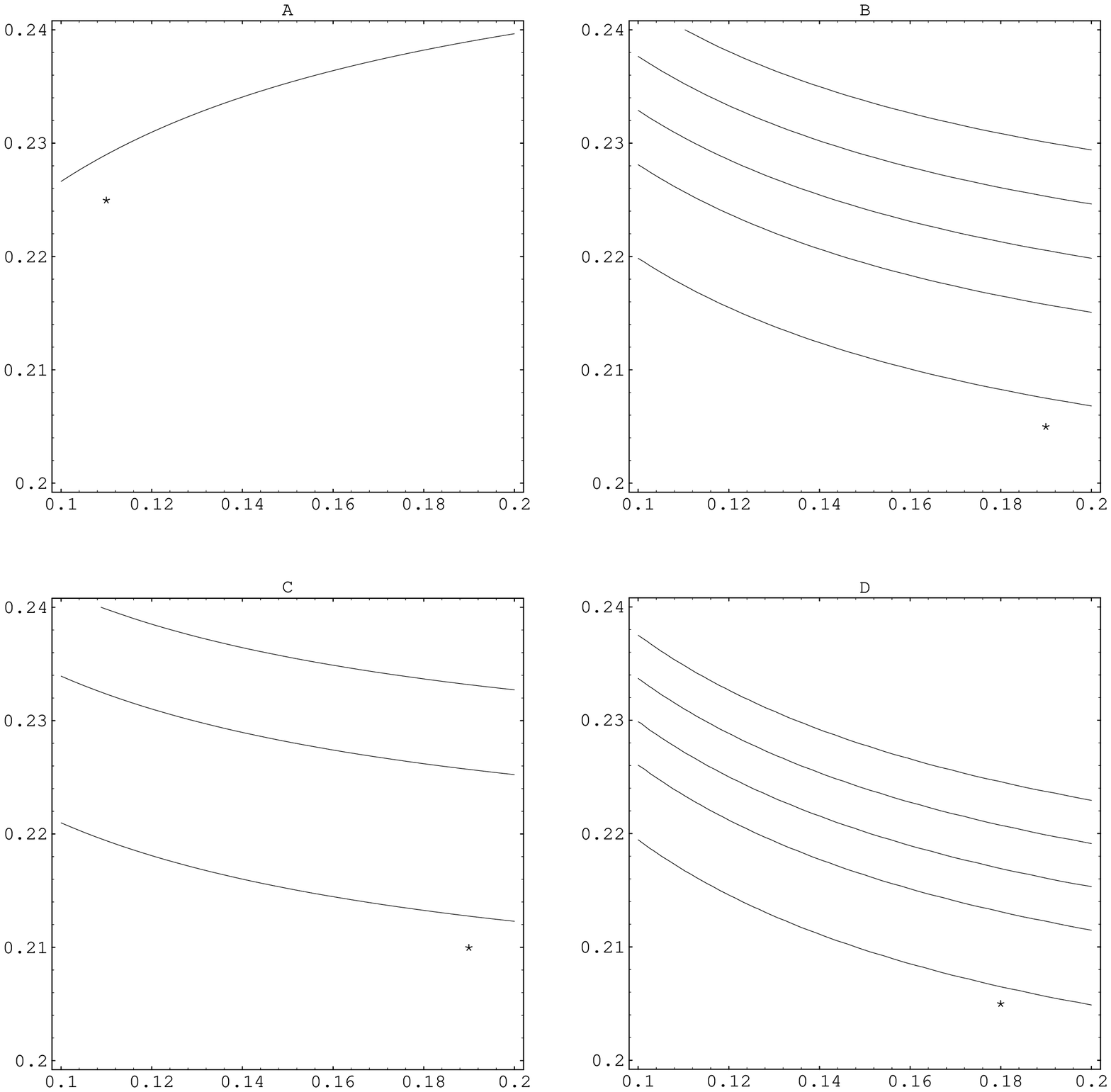}}
 {\small \it 
The figure shows the curves of level at $M_R=10^9,10^{10},
10^{11},
10^{12},5.4 \cdot 10^{13}~ GeV$ for the models of Table I.
The star indicates the value
$M^{sup}_R=5.4 \cdot10^{13}$. \\
Note that in case A only the
upper value
appears in the figure because the others are too high.}

\end{document}